\documentstyle[prb,aps,epsf,twocolumn]{revtex}
\begin{document}
\draft
\title{Van Hove Singularities in disordered multichannel quantum wires and nanotubes}  
\author{S.~H\"ugle and R.~Egger}
\address{Institut f\"ur Theoretische Physik,
Heinrich-Heine-Universit\"at, D-40225 D\"usseldorf, Germany}
\date{Date: \today}
\maketitle
\begin{abstract}
We present a theory for the van Hove singularity (VHS) in 
the tunneling density of states (TDOS) of disordered multichannel 
quantum wires, in particular multi-wall carbon
nanotubes.  We assume close-by gates which screen off electron-electron
interactions.  Diagrammatic perturbation theory within the non-crossing
approximation yields analytical expressions governing the 
disorder-induced broadening and shift of VHS's
as new subbands are opened.  This problem is nontrivial because
the (lowest-order) Born approximation breaks down close to the
VHS.   Interestingly, compared to the bulk case, the boundary TDOS shows drastically 
altered VHS, even in the clean limit.
\end{abstract}

\narrowtext

Van Hove singularities (VHS's) in the thermodynamic
density of states (DOS) have been predicted 
in 1953\cite{hove} and were observed in many experiments since then.  
The DOS for a $d$-dimensional system with dispersion relation $E(\vec k)$
can be written as an integral over the Fermi surface,
$\nu(E)=(2\pi)^{-d}\int dS /|\nabla_{\vec{k}} E(\vec{k})|$.
The quantity in the denominator is basically the group
velocity. Due to symmetries in a crystal,
the group velocity may vanish at certain momenta, 
resulting in a divergent integrand. This
divergence is integrable in three dimensions, and typically leads to a finite
DOS. In lower dimensions, however, very pronounced VHS appear.
In this paper, we focus on the one-dimensional (1D) limit, where
the VHS diverges like $\nu(E)\sim 1/\sqrt{E-E_n}$ in a clean system when
the energy $E$ approaches the threshold $E_n$ from above.  Therefore VHS's 
 appear as sharp features in the energy-dependent DOS reflecting 
the onset of new active subbands.  Similar VHS's 
exist for the tunneling density of states (TDOS) measured
at some location $x$ along the system,
\[
\nu(E,x) = \frac{{\rm Re}}{\pi\hbar}\,\int_0^\infty dt \, e^{iEt/\hbar}
 \langle \psi^{}(x,t)\psi^\dagger(x,0)\rangle ,
\]
where $\psi(x,t)$ is the electron operator.
 The TDOS is easily accessible experimentally by
measuring the conductance through a weak link or a tunnel junction. Typically,
it is obtained by means of scanning tunneling spectroscopy
(STS).\cite{sts1,sts2} In general, one has to carefully distinguish
between the DOS and the TDOS. 
While the DOS is a thermodynamic property of the
system, the TDOS is a local property and therefore  depends on
the position. In the ``bulk'' limit, the DOS and the 
(possibly coarse-grained) TDOS are expected to be identical, but
near a boundary they can strongly differ.
In the present paper we address the question: 
How is the VHS modified by disorder? 
This question is of importance for the interpretation
of experiments on multi-wall nanotubes (MWNTs).
A MWNT consists of a few concentrically arranged graphite 
shells, with a typical outermost-shell radius $R\approx$ 5 to 20
nm and lengths of up to 1 mm. 
As demonstrated by magnetoresistance 
measurements,\cite{bachtold1} electronic transport occurs only through the
outermost shell and exhibits the typical fingerprints of
diffusive behavior while single-wall nanotubes (SWNTs) are ballistic
1D quantum wires.\cite{tans} 
The observation of VHS's in recent STS experiments on SWNTs on a
metallic substrate represents a direct proof for a 1D band
structure.\cite{swnt,venema,kim} In particular, the predicted $1/\sqrt{E-E_n}$
behavior of ballistic 1D wires has been observed.

Since in MWNTs the incommensurate
inner-shell ionic potential acts on electrons in the
outermost shell, the resulting quasi-periodic potential
effectively acts as a random potential.\cite{eg2000}
We then want to understand
how VHS's develop with increasing amounts of disorder.
By intentionally damaging MWNTs, e.g.~by fast ion bombarding,
any level of disorder may be realized experimentally to test our
predictions below.  For typical MWNTs with mean free 
path $l \approx 2\pi R$, the characteristic
subband features of the TDOS should still be present, albeit considerably 
broadened and possibly shifted. 
This question is clearly also of relevance to other quasi-1D
quantum wires such as long chain molecules.
Previous experiments on intrinsically (hole-) doped MWNTs 
reported by Bachtold {\sl et al.}\cite{bacht2} have observed
power-law zero bias anomalies due to electron-electron 
interactions.\cite{eg2000}
Here we focus on a completely different scenario, where
interactions are screened off by
working on a metallic substrate which is typical for
STS measurements. Hence we treat only the non-interacting problem. 
Then spin only contributes trivial factors of two and
can be ignored.  In addition, disorder-induced scattering between the 
two distinct gapless Fermi (K) points of
the first Brillouin zone of the honeycomb lattice
is expected to be largely suppressed, and we thus
consider only one Fermi point.  Since
in typical STS experiments only the TDOS of the outer-most shell is probed,
in the following an effective single-shell model is assumed, where inner
shells only give rise to a disorder  potential for outer-shell electrons.

Nanotubes (NTs) can be thought of as graphite sheets wrapped onto a cylinder. 
The low-energy theory of a clean graphite sheet is given by the 2D Dirac Hamiltonian.
This graphite sheet is rolled up into a tube
by enforcing periodic boundary conditions
around the $y$ direction, the tube pointing along the $x$ direction.
For later convenience, we define $\beta^{-1}=\hbar v_F/2\pi R$
with $v_F=8\times 10^5$ m/s.
(To simplify notation, we often set $\hbar=v_F=1$.) 
In the absence of disorder, we have
\begin{equation}\label{h0}
H_0=\frac{1}{\beta}\sum_n\int\! \frac{dk}{2\pi}\; \psi^\dagger(\vec{k})
(\vec{\sigma}\cdot\vec{k}+M\sigma_y) \psi^{}(\vec{k}) 
\end{equation}
with $\vec{k}=(k,\omega_n)$. The ``Matsubara frequencies''
$\omega_n$ given by $E_n=2\pi n/\beta=\hbar\omega_n$ with integer $n$ arise
due to the finite radius. $\psi$ is a two-component spinor, where 
the two components reflect the sublattice degree of freedom, as the honeycomb lattice has 
a basis containing two atoms, and the Pauli matrices
$\vec{\sigma}=(\sigma_x,\sigma_y)$  
act in this space.  The ``mass'' $M$ is generally nonzero for NTs due to
chirality effects or an applied magnetic field
$B$ parallel to the tube axis.\cite{kane} For a flux $\Phi$ (in units
of the flux quantum), we get $M=2\pi\Phi/\beta$. Since 
chirality effects can always be absorbed by an adjustment of $B$,
we take $\Phi=B/B_0 $ with $B_0=h/(e \pi R^2)$.
The mass term in Eq.~(\ref{h0}) couples in the same way
as $k_y=\omega_n$, and hence we can replace
\begin{equation}\label{mn}
\omega_n\to\omega_n+2\pi\Phi/\beta=(n+B/B_0)/R
\end{equation}
in Eq.~(\ref{h0}) to include the mass term. 
For integer $\Phi$, the system is obviously not influenced
 by the magnetic field, as we can simply shift $n$ to absorb $\Phi$. 
Next we discuss our modelling of disorder.
On the one hand, there may be disorder-induced hopping events between
 nearest-neighbor sites on the honeycomb lattice
via real or virtual states provided by the inner-shell 
ionic potential. Since hopping connects different
sublattices, the resulting modulation of the hopping matrix element
leads to a random gauge field. Gauge field disorder
also captures the effects of topological defects in the outermost
shell.\cite{paco} 
However, in particular once the MWNT is intrinsically damaged, the
dominant disorder mechanism should be due to direct impurity potential 
scattering processes which are diagonal in sublattice space,
\begin{equation}\label{hv}
H_V= \int d\vec x \ V(\vec x) \psi^\dagger( \vec x)\psi (\vec x) .
\end{equation}
In what follows, we retain only disorder of the type (\ref{hv}). 
The scattering potential is taken as a static Gaussian random field with zero
mean and  
$\langle V(\vec{x}) V(\vec{x}')\rangle= \Delta_V \delta(\vec{x}-\vec{x}')$,
where $\Delta_V=\hbar^2 v_F^2\Delta/R$ defines the dimensionless disorder
strength $\Delta$.  

In addition to the bulk case, we also address the boundary TDOS arising when one 
tunnels into the end of a MWNT. In reality,
the end TDOS could of course be quite complex due to the formation of bound
states. A few lattice spacings away
from the end, however, we expect that the situation can be described 
by a continuum model.  Surprisingly, the presence of a boundary implies a drastic change
in the TDOS even in the absence of interactions, namely a
strong suppression of the TDOS close to the boundary.  Since this can be 
demonstrated already  in the absence of disorder, let us briefly discuss
the clean limit. The TDOS at position $x$ and energy $E>0$ is
$\nu(E,x)=-{\rm Im}{\rm Tr}_{\vec{k},\sigma}G(E,\vec{k},x)/\pi$,
where we have to trace over the ``spin'' and the momentum.
We consider a semi-infinite 
($x\geq 0$) tube, assuming a hard-wall potential at the boundary $x=0$.
The Greens function is then
$G_0(E,\vec{k},x)=2\sin^2(kx)/(E-\vec{\sigma}\cdot\vec{k}+i0^+)$, and hence 
\begin{eqnarray}\label{dosclean}
\frac{\nu(E,x)}{\nu_{\rm 1D}} &=& 2|E|\\
&\times& \sum_{n=-\infty}^{\infty}
\frac{\sin^2(x\sqrt{E^2-E_n^2})}{\sqrt{E^2-E_n^2}}
\Theta(E^2-E_n^2) \;,\nonumber
\end{eqnarray}
where $\Theta$ is the Heaviside step function
and $\nu_{\rm 1D}=1/\pi\hbar v_F$ serves as natural unit for the TDOS (without spin and
K point degeneracy). As one can see from Eq.~(\ref{dosclean}) with
(\ref{mn}), a gap is generally present. 
This gap varies as a function of $B$, and vanishes periodically with period $B_0$.  
Far away from the boundary,  $\sin^2(x\sqrt{E^2-E_n^2})\to 1/2$,  and
we obtain the {\sl bulk TDOS} 
(see also Ref.~\onlinecite{mint})
\begin{equation}\label{bulkdosclean}
\frac{\nu_0(E)}{\nu_{\rm 1D}}=|E|\sum_n
\frac{\Theta(E^2-E_n^2)}{\sqrt{E^2-E_n^2}} \;,
\end{equation}
which equals the thermodynamic DOS.
The well-known $1/\sqrt{E-E_n}$ VHS's
in 1D appear at the onset of new subbands.
Since the summation also includes negative values of $n$, 
a magnetic field in general causes a doubling of the VHS's, see
Eq.~(\ref{mn}). Sufficiently close to the boundary, however, 
we obtain a completely different result.
{}From Eq.~(\ref{dosclean}) we find the {\sl boundary TDOS} 
\begin{equation}\label{enddosclean}
\frac{\nu_{\rm end}(E,x)}{\nu_{\rm 1D}} =
2x^2|E|\sum_n\Theta(E^2-E_n^2)\sqrt{E^2-E_n^2} \;. 
\end{equation}
For $E\to 0$, this predicts $\nu(E)\sim E^2$ for $B=0$, in contrast to the
finite boundary TDOS for a doped tube. This behavior can be traced back to the
linear dispersion relation of Dirac fermions.
More interestingly, the typical 1D VHS of the bulk TDOS is drastically 
altered close to the boundary. Instead of a divergence, the only sign of the
opening of new subbands is a nonanalyticity at the threshold energy $E_n$,
with a {\sl square-root energy dependence of the boundary TDOS} above $E_n$.  
This possibly explaines the observation of
Ref.~\onlinecite{kim}, where no divergencies were found in the tunneling
spectrum at the end of a SWNT. 
We note that this phenomenon is quite general.
 For both Dirac and Schr\"odinger fermions,
the exponent in the TDOS energy dependence close to a VHS changes by
one when going from the bulk to the boundary 
limit.  For $E\to 0$, however, for Dirac fermions, the
exponent changes by two due to the special role of the energy $E=0$.
The relevant crossover scale between bulk and boundary behavior of
the TDOS depends on energy.  Focussing on $E$ close to but
above a given threshold energy $E_n$, this scale is
$x^* \approx \hbar v_F/\sqrt{E^2-E_n^2}$.
At zero temperature, the bulk limit is reached for $x\gg x^*$,
 and the boundary limit for
$x\ll x^*$.  For finite temperatures $T$, if the thermal scale $x_T=\hbar v_F/k_BT$ is
smaller than $x^*$, one should replace $x^*$ by $x_T$. 

We study disordered MWNTs using diagrammatic perturbation theory within the non-crossing 
approximation (NCA).\cite{abriko} The use of NCA for this problem requires some care.
On the one hand, for 2D random Dirac fermions,  diagrams with crossing impurity
lines cause the same (logarithmic) singularities as rainbow diagrams in arbitrary orders
of perturbation theory,\cite{ners} and crossing diagrams
must be treated on the same footing as the rainbow ones.
On the other hand, in the single-channel 1D limit, it is also well-known from exact
calculations that crossed diagrams are in general as important as rainbow diagrams.
Since MWNTs are in between the 1D and the 2D limit, studying
the importance of crossed diagrams is therefore mandatory.
The situation for MWNTs is fortunately quite different from both limits mentioned above.
First, there are no logarithmic singularities appearing in the self energy
expansion as in the 2D limit. Second,  evaluation of the simplest diagrams 
using standard methods \cite{abriko}
shows that rainbow diagrams are larger than crossed diagrams 
by a factor $4\pi \nu(E)/\nu_{1D}$. 
For the energy range of interest here,
we are therefore entitled to compute the TDOS within NCA.
We focus on the bulk case here and briefly comment on the boundary
result later.  

With $G_0(E,\vec{k})=(E-\vec{\sigma}\cdot\vec{k}+i0^+)^{-1}$,
the self energy to lowest order in $\Delta_V$
(Born approximation) takes the form
$\Sigma^{(1)}(E)=-i\pi\Delta_V\nu_0(E)$,
with $\nu_0(E)$ given by Eq.~(\ref{bulkdosclean}). 
Remarkably, the Born approximation breaks down in the vicinity
of a VHS. Specifically, the second-order self energy contribution obeys
\[
\left|\frac{\Sigma^{(2)}(E)}{\Sigma^{(1)}(E)}\right| =
\pi\Delta_V\partial_E\nu_0(E) \;,
\]
which diverges for $E$ approaching $E_n$
from above. We therefore must address the higher-order contributions 
to the self energy. 
Even within NCA, since the perturbation expansion is 
asymptotic, we have to arrange the order of
summation in a physically meaningful way to avoid familiar but unphysical
divergencies and inconsistencies. This would also be important for analyzing 
diagrams beyond NCA, see Ref.~\onlinecite{abriko}. 
To achieve that, we follow the 
self-consistent iterative approach proposed by Lee.\cite{lee} 
Within this approach, the self energy $\Sigma_N$ including all contributions up
to $N$th order ($N\ge 1$)
is 
$\Sigma_N(E)=\Delta_V{\rm Tr}_{\vec{k},\sigma}G_{N-1}(E,\vec{k})$,
with the corresponding Dyson equation
$G^{-1}_N(E,\vec{k})=G^{-1}_0(E,\vec{k})-\Sigma_N(E)$.
This form is then used to calculate the self energy $\Sigma_{N+1}$.
In the limit $N\to \infty$, this procedure converges leading to  
\begin{equation}\label{sigselfcon}
\Sigma(E)=\Delta_V{\rm Tr}_{\vec{k},\sigma}
\frac{1}{G^{-1}_0(E,\vec{k})-\Sigma(E)}\;,
\end{equation}
which has to be solved self-consistently for $\Sigma(E)$. One can check
explicitly in each order that Eq.~(\ref{sigselfcon}) reproduces all non-crossing
diagrams correctly. 
The result can then be used to compute the TDOS,
$\nu(E)=-\rm{Im}\ \Sigma(E)/\pi\Delta_V $.
It is not possible to simply assume an
energy-independent mean free  path for all energies. 
The energy-dependent mean free time is  $-{\rm
Im}\Sigma(E)=\hbar/2\tau(E)$, and therefore also an energy-dependent mean free
path $l(E)=v_F\tau(E)$  results.

{}From Eq.~(\ref{sigselfcon}), we obtain
($ \Sigma_R={\rm Re}\  \Sigma$)
\begin{equation}\label{sigbulk}
\Sigma(E)=-i\Delta\frac{\hbar v_F}{R}
\sum_n\frac{(E-\Sigma(E))\;{\rm
sgn}(E-\Sigma_R)}{\sqrt{(E-\Sigma(E))^2-E_n^2}} \;,  
\end{equation}
where correct units were restored for clarity. Equation (\ref{sigbulk}) can
easily be solved numerically.  
Once we know the self energy, the TDOS follows as
\begin{equation}\label{dosbulk}
\frac{\nu(E)}{\nu_{\rm 1D}}={\rm Re}
\sum_n\frac{(E-\Sigma(E))\;{\rm
sgn}(E-\Sigma_R)}{\sqrt{(E-\Sigma(E))^2 -E_n^2}}\;. 
\end{equation}
For $\Delta>0$, the finite imaginary part of $\Sigma$ in the denominator of 
(\ref{dosbulk}) causes a broadening of the VHS, whereas
the real part causes an energy shift of the peaks. 
For the numerical evaluation of Eq.~(\ref{sigbulk}), a cutoff for the band
index $n$ has to be used, which is naturally given from the bandstructure.  
For instance, for armchair NTs, the number of subbands is
given by $2N=8\pi R/\sqrt{3}a$.\cite{dressel} For
$R=10$~nm, we get $N\approx 295$.
The results are, however, not very sensitive to the
precise choice of this cutoff. 
Equations (\ref{sigbulk}) and (\ref{dosbulk}) can be used to fit
experimental data for the TDOS of MWNTs. Assuming that $R$ is known, since
$\Delta$ is the only fit parameter, the disorder strength can 
then be determined directly from the TDOS which should allow for detailed
comparison to STS experiments on MWNTs.  
This would provide precious information on the disorder strength.

\begin{figure}
\epsfxsize=0.9\columnwidth
\epsffile{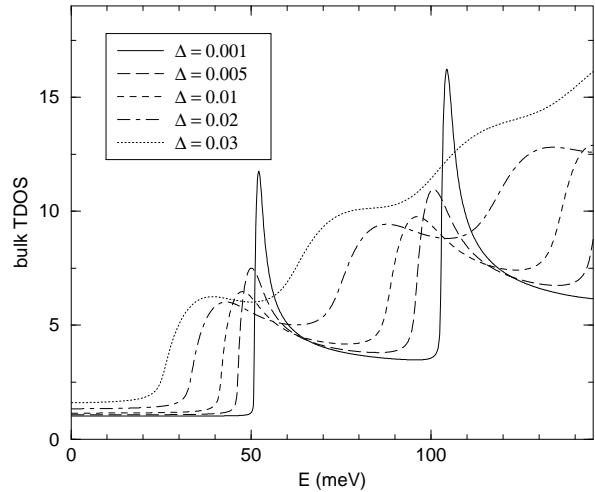}
\caption[]{\label{fig1} 
Bulk TDOS of a disordered tube with $R=10$~nm at $B=0$ for different values
of $\Delta$. The broadening of the VHS with increasing $\Delta$ is clearly visible,
 as well as their shift towards smaller energies.
}   
\end{figure}

\begin{figure}
\epsfxsize=0.9\columnwidth
\epsffile{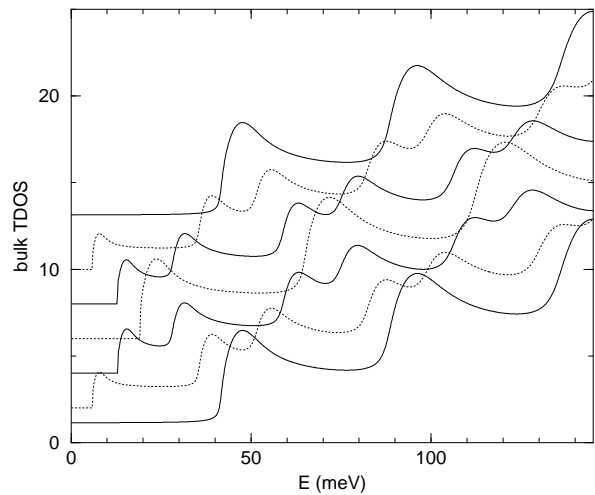}
\caption[]{\label{fig2} 
Bulk TDOS for $\Delta=0.01, R=10$~nm and different values of $B$. The
curves corresponding to different $B$ are shifted vertically by
the same amount for better visibility. The lowest curve corresponds to $B=0$,
 then $B$ increases in steps of $\Delta B=2.2$~T. Notice
the opening and closing of a gap with period $B_0=13.2$~T. 
}   
\end{figure}

Figure \ref{fig1} shows the strong broadening of the VHS due to disorder. For
sufficiently strong disorder, namely when motion around the
circumference becomes diffusive, the peaks can even disappear completely
above a certain energy which decreases with increasing $\Delta$. 
In the region where no VHS's are present, 
the TDOS apparently behaves like a power law, $\nu(E)\sim E^\alpha$,  
with $\alpha(\Delta)\le 1$
 This power law scaling holds remarkably well for
$\Delta\le 0.05$. Note that energies are not measured 
 relative to $E_F$, and therefore this power law is unrelated
to the findings of Ref.~\onlinecite{bacht2}.
Also it has a different origin than the power law 
found in the 2D case that rests upon the inclusion of crossed 
diagrams.\cite{ners} 
In addition, the  positions of the VHS's are
 shifted to smaller energies with increasing $\Delta$.
This shift grows approximately linearly
 with $\Delta$, and the relative shift,
compared to the position of the VHS in the clean system, can be
up to 20\% depending on the disorder strength. 
Since the radius of a NT 
is often determined from the relative positions of the VHS,
this observation suggests that such interpretations need
to be taken with some caution.
The disorder-induced shift has to be taken into account to
obtain correct results. With decreasing radius and for increasing order $n$ of
the VHS, the relative shift becomes systematically smaller.
Next, if we consider a fixed $\Delta$ and vary a magnetic field $B$
applied parallel to the tube axis, we find the 
situation depicted in Fig.~\ref{fig2}. Again, there is a doubling of the 
VHS's, and the shift of the positions
varies periodically with $B$. As in the clean case, there is a
gap even in the presence of disorder. But due to the $\Delta$-dependent shift
of the VHS towards smaller energies, 
the gap gets partially filled with states
and is therefore smaller than in the clean case. 
However, magnetic field effects are only weakly affected by disorder. 
Finally, we briefly turn  to the boundary limit. 
Since it is easy to generalize 
Eqs.~(\ref{sigbulk}) and (\ref{dosbulk}) to obtain the 
boundary TDOS, we only mention the modifications compared to the
clean case. The disordered boundary TDOS is suppressed compared to the clean
case, where the suppression increases with increasing $\Delta$. The
positions of the van Hove nonanalyticities at the opening of new subbands are
shifted to {\sl higher} energies, although the shift in the bulk case was to 
smaller energies. The form of the  nonanalyticity is not significantly changed
by disorder, but approximately retains the  square-root energy dependence.

To conclude, we have calculated the TDOS of disordered multichannel
quantum wires, with special emphasis on MWNTs.  In the present theory, 
electron-electron interactions are supposed to be screened off by a metallic substrate
or a close-by gate. Focussing on static disorder, 
within NCA a self-consistent non-pertubative
summation of all diagrams for the self energy yields 
an analytical result for the disorder-broadened VHS's.
For given radius of the MWNT, this result involves only one
parameter (the disorder strength $\Delta$), which should allow
for a detailed comparison to STS
experiments on MWNTs, where the disorder strength can
be tuned at will.

We thank A.~O.~Gogolin  for useful discussions. Financial support by the DFG 
under the Gerhard-Hess program is acknowledged.

\end{document}